\documentstyle[epsfig]{europhys}

\def\etal{{\hbox{{\tenit\ et al.\/}\tenrm :\ }}}

\newif\ifboo \boofalse

\def\Review#1{\boofalse{\it #1},}

\def\Vol#1{\ifboo Vol. {\bf #1}\else{\bf #1}\fi}
\def\Year#1{\ifboo #1\else(#1)\fi}

\def\Page#1{\ifboo {\rm p. #1}\else{\rm #1}\fi}

\begin{document}
\euro{}{}{}{2005}
\Date{}
\shorttitle{T. VERECHTCHAGUINA \etal: FIRST PASSAGE TIME DENSITIES}
\title{First Passage Time Densities in Non-Markovian Models with Subthreshold
Oscillations}
\author{T. Verechtchaguina, I.M. Sokolov, and L. Schimansky-Geier}
\institute{Institute for Physics, Humboldt-University at Berlin, Newton
Str. 15, D-12489 Berlin, Germany}
\rec{September 2005}{in final form}
\pacs{\Pacs{05}{40$-$a}{Fluctuation phenomena, random processes, noise,
and Brownian motion}
\Pacs{02}{50.Ey}{Stochastic processes}
}
\maketitle

\begin{abstract}
  Motivated by the dynamics of resonant neurons we consider a
  differentiable, non-Markovian random process $x(t)$ and particularly the time
  after which it will reach a certain level $x_b$. The probability density of
  this first passage time is expressed as infinite series of
  integrals over joint probability densities of $x$ and its velocity
  $\dot{x}$.  Approximating higher order terms of this series through
  the lower order ones leads to closed expressions in the cases of
  vanishing and moderate correlations between subsequent crossings of
  $x_b$.  For a linear oscillator driven by white or coloured Gaussian
  noise, which models a resonant neuron, we show that these
  approximations reproduce the complex structures of the first passage time
  densities characteristic for the underdamped dynamics,
  where Markovian approximations (giving monotonous first passage
  time distribution) fail.
\end{abstract}

\section{Introduction}
The first passage time (FPT) of a stochastic process $x(t)$ starting
at $t=0$ from a given initial value within an \textit{a priory}
prescribed domain $\Omega$ of its state space is the time $T$ when
$x(t)$ leaves $\Omega$ for the first time. This concept was originally
introduced by E.  Schr\"{o}dinger when discussing behaviour of Brownian
particles in external fields \cite{S1915}. A large variety of problems
ranging from noise in vacuum tubes, chemical reactions and nucleation
\cite{HTB1990} to stochastic resonance \cite{LBM1991}, behaviour of
neurons \cite{T1988}, and risk management in finance \cite{R2001} can
be reduced to FPT problems. 

We will assign ${\cal F}(T)$ to be the flux through the boundary of
$\Omega$ at time $T$, \textit{i.e.} the probability density of the first passage
time. Approaches to find ${\cal F}(T)$ are typically
based either on the Fokker-Planck equation with an absorbing
boundary \cite{R1989} or on the renewal theory \cite{vK1992}. Despite the
long history, explicit expressions for the FPT density are known only
for a few cases. These include overdamped particles in the force free
case, with time-independent constant forces and linear forces under
influence of white noise \cite{T1988,TM1977,GM1964,S1967}, as well as
the case of a constant force under coloured noise \cite{B2004}.
Reasonable approximations exist for a few nonlinear forces
\cite{SH1989,LFS+}.

If the relaxation time $\tau_{rel}$ of the system to a quasistationary
distribution in $\Omega$ is much smaller than the typical first
passage time, the escape from $\Omega$ occurs from this
quasiequilibrium state independent of the initial condition. Escapes
occur with a constant \textit{rate} inversely proportional to the mean
FPT. This is the case for many chemical reactions and nucleation
processes \cite{HTB1990} as well as for the leaky integrate-and-fire
model of a neuron \cite{report}. On times $T$ exceeding $\tau_{rel}$ the
FPT probability density ${\cal F}(T)$ decays exponentially.

If the scale separation between relaxation and escape times does not
hold, the dependence on the initial conditions gets crucial \cite{SSL+2001}.
This situation is found in resonant neurons
\cite{EKH+2004,VSS2004,I2001,LSS2002,Brunel} where $x$, here the
voltage variable, exhibits damped subthreshold oscillations around the
rest state being the attractor for deterministic dynamics of the
system. If $x(t)$ exceeds the excitation threshold $x_b$, a new spike
begins. After spiking the voltage is reset to a fixed value $x_0$ far
from the rest state and can reach $x_b$ again \emph{prior} to
relaxation.  ${\cal F}(T)$ giving the probability density function (PDF)
of intervals between two consecutive spikes strongly deviates from an exponential.  This is
the situation we have in mind when developing the approach for
obtaining ${\cal F}(T)$ for a non-Markovian differentiable random
processes $x(t)$ starting for $t=0$ at $x(0)=x_{0}<x_{b}$.

Our approach is based on a series expansion for ${\cal F}(T)$ which is
known for several decades \cite{F1980}.  But it was never used
for explicit calculations.  The approach is based on the theory of
level crossings originally put forward by S.O. Rice \cite{R1945}. A
generalisation of his approach (based on what is called the
Wiener-Rice series \cite{S1951}) was used by Stratonovich to estimate
the mean time spent by a stochastic process above the given level
$x_{b}$ \cite{S1967}.  The exact expression for ${\cal F}(T)$ is
analogous to the Wiener-Rice series, but corresponds to the case when
the initial value $x(0)$ differs from the threshold $x_{b}$.
 
We first give an elementary derivation of this result and show then
how this series can be used to obtain analytical expressions based
on decoupling approximations.  Explicit calculations are performed for
an underdamped harmonic oscillator driven by white or coloured Gaussian
noise, \textit{i.e.}  for a resonate-and-fire neuron with subthreshold
oscillations and reset.  The linearity of the model simplifies
calculations but is not crucial for the approach.

\section{Level crossings and first passage times}
Let us first discuss the probability $n_1(x_b,t|\{x_0\},0)dt$ that a
differentiable random process $x(t)$ starting from a fully defined
initial condition $\{x_0\}=\{x(0),\dot{x}(0),...\}$ (corresponding to
its Markovian embedding) with $x(0)<x_b$ crosses the level $x_{b}$
between $t$ and $t+dt$ with positive velocity $v(t)=\dot{x}(t)>0$
(\textit{i.e.} performs an \textit{upcrossing}).  The upcrossing can only occur
for such $x(t)$ that $x_{b}-vdt<x(t)<x_{b}$.  The probability of this
is
$\int_{x_{b}-vdt}^{x_{b}}P(x,v,t|\{x_0\},0)dx=|v|P(x_b,v,t|\{x_0\},0)dt$,
where $P(x,v,t|\{x_{0}\},0)$ is the joint PDF for $x$ and $v$ under
given initial conditions. Integration over all $v>0$ then gives
$n_{1}(x_{b},t|\{x_0\},0)=\int_{0}^{\infty}vP(x_{b},v,t|\{x_0\},0)dv$.
The joint probabilities for multiple upcrossings $n_p(t_p,\dots,
t_1)dt_p\dots dt_{1}$ in each of $p$ time intervals
$(t_{1},t_{1}+dt_{1}),\dots (t_{p},t_{p}+dt_{p})$ follow in the same
way:
\begin{equation}
n_{p}(t_{p},\dots,t_{1})=
\int _{0}^{\infty }dv_{p}\dots \int_{0}^{\infty }dv_{1}
v_{p}\dots v_{1}P(x_{b},v_{p},t_{p};\dots ;x_{b},v_{1},t_{1}|\{x_{0}\},0).
\label{E:RiceJoint}
\end{equation}
The initial conditions and $x_b$ will be omitted in what follows. 

The function ${\cal F}(T)$ is given by the fraction of all
trajectories $x(t)$ which perform \textit{the first} upcrossing of
$x_{b}$ at time $T$. All such trajectories are accounted for in
$n_{1}(T)$. However, $n_{1}(T)$ also considers trajectories which
might have an earlier upcrossing at some $0<t_{1}<T$. Since they
should not contribute to ${\cal F}(T)$, their fraction should be
subtracted from $n_{1}(T)$ by taking
$n_{1}(T)-\int_{0}^{T}n_{2}(T,t_{1})dt_{1}$.  This excludes all
trajectories which cross $x_b$ exactly twice until $T$.  However the
trajectories crossing $x_b$ three times, i.e at $T$ and at two earlier
moments $t_i<T,(i=1,2)$, are not correctly accounted.  Each such
trajectory is added once in $n_1(T)$, but subtracted two times in
$\int_{0}^{T}n_{2}(T,t_{1})dt_{1}$, since this term accounts for the
pairs of upcrossings at $(T,t_i)$ with $i=1,2$. To account for this we
have to add once the amount of trajectories with three upcrossings
again:
$n_{1}(T)-\int_{0}^{T}n_{2}(T,t_{1})dt_{1}+\frac{1}{2!}\int_{0}^{T}
\int_{0}^{T}n_{3}(T,t_{2},t_{1})dt_{2}dt_{1}$ (the factor $1/2!$ in
the last term accounts for permutations of $t_{i}$).

Generally, if a trajectory crosses $x_b$ at time $T$ and at $N$
earlier times $t_i<T$, $i=1 \dots N$, then in $\frac{1}{p!}\int \dots
\int n_{p+1}(T,t_p,\dots,t_1)dt_p\dots dt_1$ it is accounted for
exactly $C_N^p$ times ($C_N^p$ stands for the number of combinations).
Note, that $\sum_{p=0}^{N}(-1)^{p}C_N^p=(1-1)^N=0$.  Thus in the
alternating sum of this kind containing $N+1$ terms, all trajectories
crossing $x_b$ at time $T$ and having $1,2,\dots N$ additional
upcrossings are excluded. Extending the sum to infinity we exclude all
superfluous trajectories. Only trajectories remain for which the
upcrossing at time $T$ was the first one. Thus the expression for the
first passage time probability density reads:
\begin{equation}
{\cal F}(T)=\sum_{p=0}^{\infty }\frac{(-1)^{p}}{p!}
\int_{0}^{T}dt_{p}\dots
\int_{0}^{T}dt_{1}n_{p+1}(T,t_{p},\dots ,t_{1}).
\label{E:FPTgeneral}
\end{equation}

In Ref. \cite{vK1992,S1967} it was shown that ${\cal F}(T)$ can be
equivalently expressed in terms of the correlation functions between
upcrossings (cross-cumulants) $g_{p}(t_{p},\dots , t_{1})$.  They are
related to the joint densities $n_p(t_p,\dots ,t_1)$ via
\begin{eqnarray}
&& g_1(t_1) =n_1(t_1), \nonumber \\
&& g_2(t_2,t_1) =n_2(t_2,t_1)-n_1(t_1)n_1(t_2), \\ \nonumber
&& g_3(t_3,t_2,t_1) = n_3(t_3,t_2,t_1)-3\{ n_1(t_1)n_2(t_3,t_2)\}_s+
2n_1(t_1)n_1(t_2)n_1(t_3), \qquad \dots .
\label{E:CorrFuncDens}
\end{eqnarray}
Here $\{\dots\}_s$ denotes the symmetrisation of the expression in the
brackets with respect to all permutations of its arguments.  The
expression for ${\cal F}(T)$ in terms of correlation functions reads
\cite{S1967}:
\begin{equation}
{\cal F}(T) = S^{\prime}(T)e^{-S(T)} \label{E:GenForm}
\end{equation}
with
\begin{equation}
S(T)= - \sum_{p=1}^{\infty }\frac{(-1)^{p}}{p!}
\int_{0}^{T}dt_{p}\dots \int_{0}^{T}dt_{1}g_{p}(t_{p},\dots ,
t_{1}).\label{E:Sexact}
\end{equation}
The function $S^{\prime }(T)$ can be interpreted as a the
time-dependent escape rate.

We emphasise that expressions eqs.(\ref{E:FPTgeneral}),(\ref{E:GenForm}) and
(\ref{E:Sexact}) do hold for all differentiable random processes (ones whose
velocity $v(t)$ is defined at any time). 

\section{Decoupling approximations for the FPT density}
Dealing with infinite series useful approximations can either be based
on the truncation of the series after several first terms calculated
exactly, or on expressing the higher order terms approximately through
the lower order ones what might lead to a closed analytical form.
Truncation approximations for eq.(\ref{E:FPTgeneral}) are not
normalised and hold on short time scales only diverging at longer
times (due to the miscount of trajectories with several upcrossings).
We discuss here the approximations of the second type which
are normalised and can be used in the whole time domain.  Each such
approximation is based on a subsummation in eq.(\ref{E:Sexact}) for
$S(T)$. Note, only expressions guaranteeing positive rates $S(T)$ are
allowed.
 
The simplest approximation is based on neglecting all terms in
eq.(\ref{E:Sexact}) except for the first one. It leads to
\begin{equation}
S(T)=\int_0^T n_1(t)dt,  \label{E:Hertz}
\end{equation}
where $n_{1}(t)$ is given by eq.(\ref{E:RiceJoint}) with $p=1$.
This corresponds to neglecting all correlations between upcrossings of the level $x_{b}$,
\textit{i.e.} to factorisation of $n_{p+1}(T,t_p,\dots,t_1)$ into a product of
one-point densities $n_1(T)n_1(t_p) \dots n_1(t_1)$ in eq.(\ref{E:FPTgeneral}).
Then the series, eq.(\ref{E:FPTgeneral}) sums up into ${\cal F}(T)\approx n_1(T)\exp \left(
-\int_0^T n_1(t)dt\right)$, which is equivalent to eq.(\ref{E:Hertz}).
The approximation will be refereed to as a \textit{Hertz approximation} since 
the form of ${\cal F}(T)$ resembles the Hertz distribution \cite{H1909}.

The second order approximation expresses all higher order correlation
functions through the first and the second ones. It was used by
Stratonovich \cite{S1967} in the context of peak duration. The first
and the second correlation functions are taken exactly, and the higher
ones are approximated by the combinations of these two. For $p \geq 2$
one thus has
\begin{equation}
g_p(t_p,\dots,t_1) \approx  (-1)^{p-1}(p-1)! n_1(t_p) \dots n_1(t_1)
\{ R(t_1,t_2)R(t_1,t_3) \dots R(t_1,t_p) \}_s
\label{E:CorrFuncApp}
\end{equation}
with the correlation coefficient
\begin{equation}
R(t_{i},t_{j})=1-\frac{n_2(t_{i},t_{j})}{n_{1}(t_{i})n_{1}(t_{j})}.
\label{E:CorCoef}
\end{equation}
The correlation coefficient $R(t_{1},t_{2})$ is equal to unity for
$t_{1}=t_{2}$ and vanishes for large values of $|t_{1}-t_{2}|$.
Substitution of eq.(\ref{E:CorrFuncApp}) into eq.(\ref{E:Sexact})
delivers then the \textit{Stratonovich approximation} for ${\cal
  F}(T)$ with the time-dependent escape rate being
\begin{equation}
S(T)=-\int_{0}^{T}n_{1}(t)\frac{\ln \left[
1-\int_{0}^{T}R(t,t^{\prime })n_{1}(t^{\prime })dt^{\prime }\right] }
{\int_{0}^{T}R(t,t^{\prime })n_{1}(t^{\prime })dt^{\prime }}dt.
\label{E:Straton}
\end{equation}

\section{Noise driven harmonic oscillator}
Let us now illustrate the method by applying it to the coordinate
$x(t)$ of a harmonic oscillator driven by a Gaussian noise $\eta(t)$
\begin{equation}
\dot{x} =v, \quad  \dot{v} =-\gamma v-\omega _{0}^{2}x+\eta (t). \label{E:HarmOscil}
\end{equation}
Two cases will be considered: (i) white noise, $\eta (t)=\sqrt{2D}\xi
(t)$, and (ii) Ornstein-Uhlenbeck noise, $\dot{\eta}=-\tau ^{-1}\eta
+\sqrt{2D}\tau ^{-1}\xi (t)$, with $\xi (t)$ being white Gaussian
noise of intensity $1$.  Eq.(\ref{E:HarmOscil}) with boundary at
$x_b$ and reset at $x(0)=x_{0},\dot{x}(0)=v_{0}$ is relevant for a
modelling of neuronal dynamics.  In the underdamped regime $\gamma < 2
\omega_0$ it describes an excitable dynamics with damped subthreshold
oscillations characteristic for resonant neurons
\cite{EKH+2004,VSS2004,I2001,LSS2002,Brunel}. In the overdamped regime
$\gamma>2\omega_0$ it describes the behaviour of nonresonant neurons
\cite{EKH+2004,J1994}.  In our calculations we fix $\omega _{0}=1$,
take initial conditions to be $x_{0}=-1,v_{0}=0$, and set the
absorbing boundary at the threshold $x_{b}=1$.

Our basic model has an advantage that all transition probability
densities in eq.(\ref{E:RiceJoint}) are Gaussian and can be expressed
through the correlation functions of their arguments, which are easily
obtained using the spectral representation \cite{R1989}.  Then
$n_1(T)$ is obtained in closed analytical form; the joint densities of
multiple upcrossings are readily obtained through numerical evaluation
of integrals in eq.(\ref{E:RiceJoint}).

The Hertz and the Stratonovich approximations eqs.(\ref{E:Hertz}) and
(\ref{E:Straton}) hold if all correlations decay considerably within
the typical time interval between upcrossings. The decay of
correlations is described by the relaxation time $\tau_{rel}
=2/\gamma$ of the process.  The mean interval between two successive
upcrossings $T_R$ is the inverse stationary frequency of upcrossings
$1/T_R=n_0 = \lim_{t \rightarrow \infty}n_1(t)$ \cite{R1945}. $n_0$ is
known as the Rice frequency and is given by
\begin{equation}  \label{E:RiceFrequency}
n_0=(2\pi)^{-1} \sqrt{ r^{\prime \prime}_{xx}(0) / r_{xx}(0)}
e^{-x_{b}^{2}/2r_{xx}(0)}
\end{equation}
where $r_{xx}(t)=\langle x(t)x(0) \rangle$ is the correlation function
of the process. In our case $T_R=2 \pi \omega_0^{-1} e^{\gamma x_b^2
  \omega_0^2/2D}$. Therefore, the Hertz approximation holds for
$\tau_{rel} \ll T_R$; for the Stratonovich approximation this
condition can be weakened to $\tau_{rel} < T_R$ arising from the
condition that the argument of the logarithm in eq.(\ref{E:Straton})
is positive.

\begin{figure}[tbp]
\begin{center}
\epsfig{file=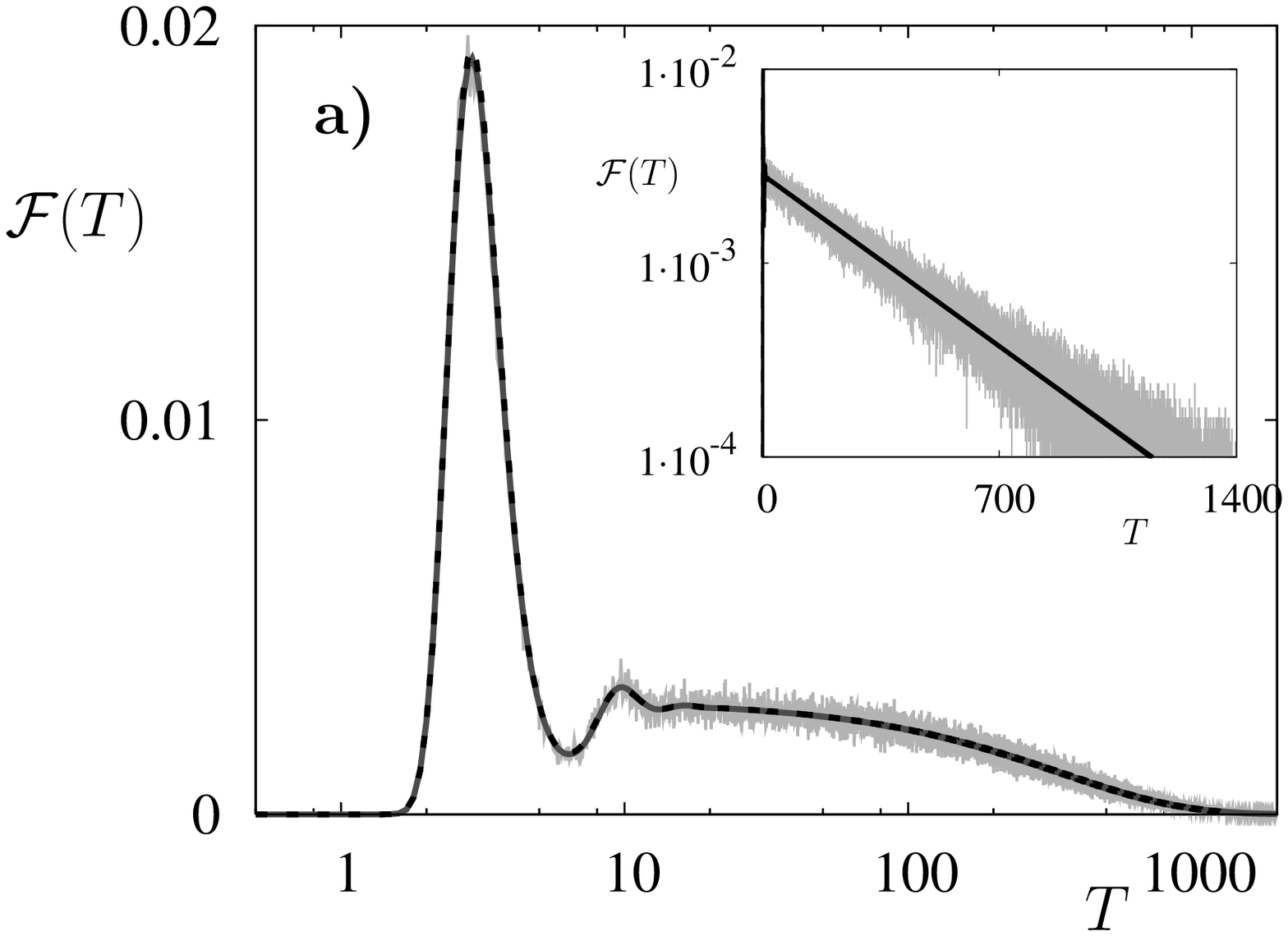,width=6.5cm} \epsfig{file=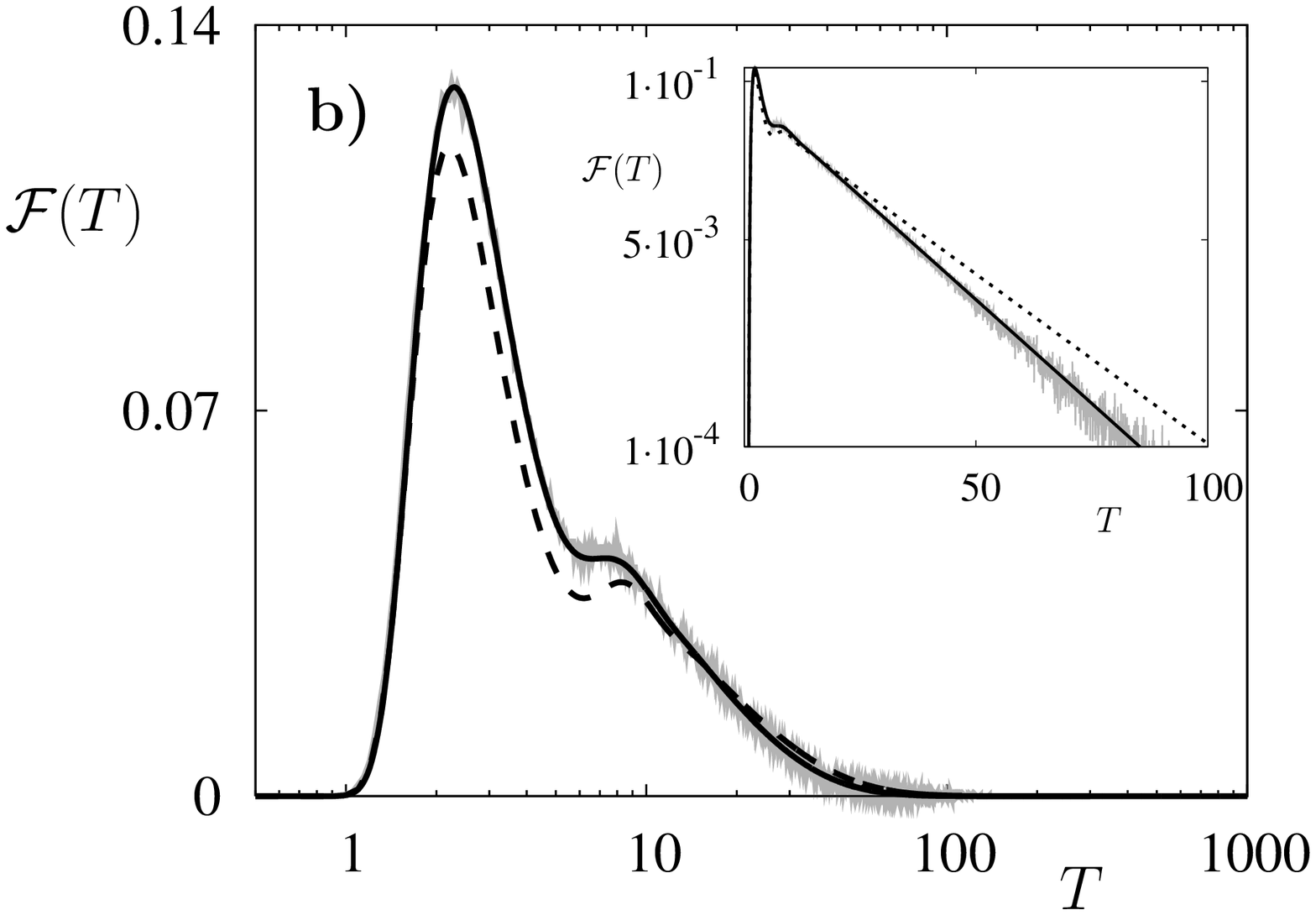,width=6.5cm}
\epsfig{file=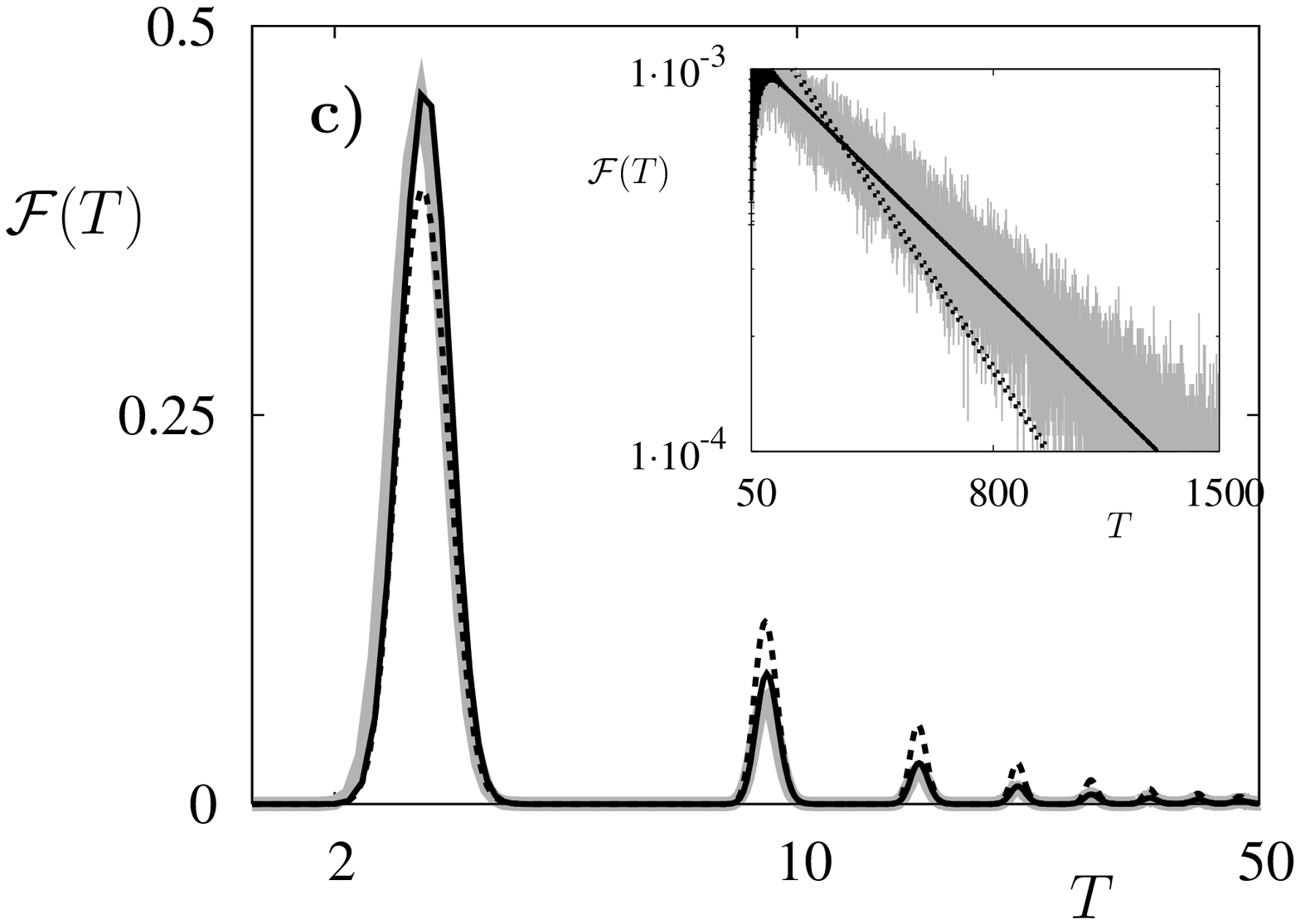,width=6.5cm} \epsfig{file=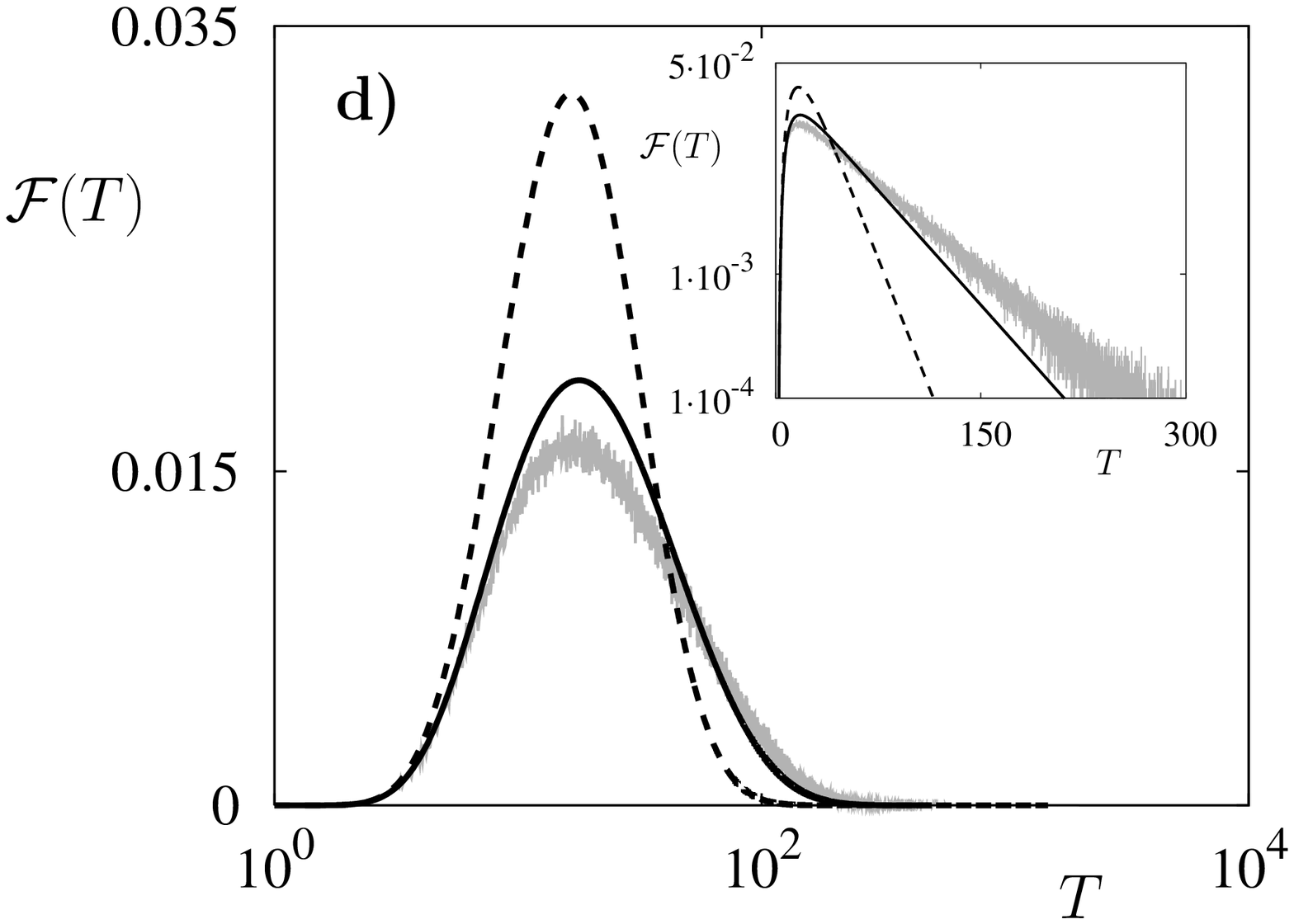,width=6.5cm}
\end{center}
\caption{FPT probability density for harmonic oscillator driven by Gaussian
white noise. Simulation results are shown with a gray line, Hertz
approximation with a black dashed line, and Stratonovich approximation with
a black solid line. Note the logarithmic scale in $T$. The insets show the
same curves on the logarithmic scale in ${\cal F}(T)$. The parameters
are: $\omega_0=1,x_0=-1,v_0=0,x_b=1$, a) $\gamma = 0.8, D=0.1,
\tau_{rel}=2.5, T_R=343$, b) $\gamma=0.8, D=0.44, \tau_{rel}=2.5, T_R=15.6$,
c) $\gamma=0.08, D=0.01, \tau_{rel}=25, T_R=343$, d) $\gamma=10.0, D=5.5,
\omega_0/\gamma=0.1$}
\label{F:FPTunderdamped}
\end{figure}

Let us first concentrate on the case (i) with white noise. In fig.
\ref{F:FPTunderdamped} the FPT probability density obtained from
simulations is compared with the Hertz and Stratonovich approximations
(eq.(\ref{E:Hertz}) and eq.(\ref{E:Straton})).  The probability to
reach $x_b$ is higher in the maxima of the subthreshold oscillations.
The initial phase of these oscillations is fixed by initial
conditions.  Thus on shorter time scales ${\cal F}(T)$ shows the
multiple peaks following with the frequency of damped oscillations
$\sqrt{\omega_0^2-\gamma^2/4}$. On long times $T\gg \tau_{rel}$ the
quasiequilibrium establishes and FPT densities decay exponentially
(see insets in fig.  \ref{F:FPTunderdamped}). The number of visible peaks
depends on the relation between $\tau_{rel}$ and the period of
oscillations and is given by the number of periods elapsing before the
quasiequilibrium is achieved.

In fig.\ref{F:FPTunderdamped}(a) the parameters are chosen to be
$\gamma=0.8,D=0.1$, corresponding to moderate friction and moderate
noise intensity. For given parameter values $\tau_{rel}=2.5$ and
$T_R=343$, so that $\tau_{rel} \ll T_R$, both Hertz and Stratonovich
approximations hold and reproduce well the FPT density in the whole
time domain.

In the case of moderate friction and stronger noise the upcrossings become
more frequent and $T_R$ decreases. The FPT changes its form to practically
monomodal with the only maximum and a small shoulder separating it from the
tail. An example is given in fig.\ref{F:FPTunderdamped}(b) with
$\gamma=0.8,D=0.44$ which correspond to $\tau_{rel}=2.5$ and $T_R=15.6$. The
Stratonovich approximation complies very well with the simulations, while
the Hertz approximation fails to reproduce the details of the distribution:
It underestimates ${\cal F}(T)$ on short times, and shows slower
exponential decay in the tail than the one observed in simulations (see the
inset).

Finally, for small friction and low noise the upcrossings are rare, but
the relaxation time is large. The FPT probability density exhibits multiple
decaying peaks. In fig.\ref{F:FPTunderdamped}(c) $\gamma=0.08,D=0.01
$ corresponding to $\tau_{rel}=25, T_R=343$. Again, the Stratonovich
approximation performs well, while the Hertz approximation underestimates
the first peak, overestimates all further peaks and decays in the tail
faster than the simulated FPT density.

Fig. \ref{F:FPTunderdamped}(d) corresponds to the overdamped regime
($\gamma>2\omega_0$) where the parameters are chosen to be
$\gamma=10.0,D=5.5$ corresponding to $\omega_0/\gamma = 0.1$.  The
condition $\tau_{rel}<T_R$ is always fulfilled in the overdamped case.
However, with increasing friction the process $x(t)$ approaches the
Markovian (diffusion) one, for which the pattern of upcrossings is
very inhomogeneous.  The successive crossings are strongly clustered
\cite{S1967}.  The upcrossings within a cluster are not independent
even if their mean density $n_0$ is low.  This limits the accuracy of
our approximations: The Stratonovich approximation starts to be
inaccurate, and the Hertz approximation fails.

For $T$ large ${\cal F}(T)$ decays exponentially, ${\cal F}(T) \propto
\exp(- \kappa T)$.  The decrement of this decay is obtained from long
time asymptotics: $\kappa T=\lim_{T \rightarrow \infty} S(T)$. Thus,
in the Hertz approximation eq.(\ref{E:Hertz}) one gets
$\kappa=(1/T)\lim_{T \rightarrow \infty} \int_0^T n_1(t) dt =
n_0T/T=n_0 $.  The behaviour in the Stratonovich approximation
eq.(\ref{E:Straton}) is determined by $\lim_{t,t^{\prime} \rightarrow
  \infty} \int_0^{T} R(t,t^{\prime})n_1(t^{\prime})dt^{\prime} \approx
n_0 \tau_{cor}$ with $\tau_{cor}$ given by $\tau_{cor}= \lim
\limits_{t \rightarrow \infty} \int_0^{\infty}R(t,t^{\prime})
dt^{\prime}$. Note that $\tau_{cor}$ is not necessary positive.
Inserting this expression into eq.(\ref{E:Straton}) and expanding the
logarithm up to the second term we get $\kappa =
n_0(1+\frac{1}{2}n_0\tau_{cor})$ providing the second order correction
to the previous expression.  The value of $\tau_{cor}$ for the
parameter set as in fig. \ref{F:FPTunderdamped}(a) is
$\tau_{cor}=-1\cdot10^{-3}$, for parameters as in fig.
\ref{F:FPTunderdamped}(b) $\tau_{cor}=5.25$, and for parameters as in
fig. \ref{F:FPTunderdamped}(c) $\tau_{cor}=-396.7$. The long time
asymptotics obtained with these $\tau_{cor}$ values reproduce fairly
well the decay patterns found numerically.

\begin{figure}[tbh]
\begin{center}
\epsfig{file=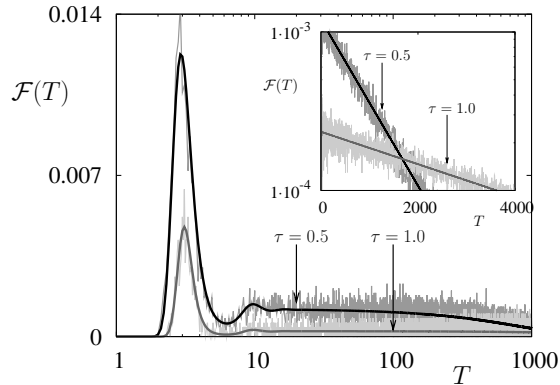,width=7.5cm}
\end{center}
\caption{Simulation results and the Hertz approximation for the FPT
probability density for harmonic oscillator driven by Ornstein-Uhlenbeck
noise with correlation times $\tau=0.5$ (upper curves) and $\tau=1.0$ (lower
curves) and other parameters as in fig. \ref{F:FPTunderdamped}(a). Note the
logarithmic scale in $T$. The inset shows the same curves on the logarithmic
scale in $ {\cal F}(T)$.}
\label{F:FPTOU}
\end{figure}

Finally we consider our model (ii) having a higher dimension of its state
space. In fig. \ref{F:FPTOU} we show the simulated FPT probability density
and the Hertz approximation for the system eq.(\ref{E:HarmOscil}) driven by
the Ornstein-Uhlenbeck noise for two different values of the correlation
time $\tau=0.5$ and $\tau=1.0$. In both cases the Hertz approximation is 
absolutely sufficient.

\section{Conclusions}
For the case of moderate friction and moderate noise intensity the
Hertz approximation is absolutely sufficient. The Stratonovich
approximation performs evenly well and does not lose accuracy for high
noise intensity. The validity region of approximations covers all
different types of subthreshold dynamics, and reproduces all
qualitative different structures of the FPT PDF: from monomodal
through bimodal to multimodal densities with decaying peaks. The
approximations work for the systems of whatever dimension and are
especially effective for the processes with narrow spectral density,
exactly when Markovian approximations fail.

We acknowledge financial support from the DFG by Graduierten-Kolleg
268 and the Bernstein Center for Computational Neuroscience, Berlin
and Sfb555.

\end{document}